\documentclass[]{spie}  

\usepackage{amsmath,amsfonts,amssymb}
\usepackage{graphicx}
\usepackage[colorlinks=true, allcolors=blue]{hyperref}
\usepackage[normalem]{ulem}
\usepackage{multirow}

\title{Antenna characterization for the HIRAX experiment}

\author[a]{Emily R. Kuhn}

\author[b,a]{Benjamin R.~B. Saliwanchik}

\author[c]{Kevin Bandura}
\author[d]{Michele Bianco}
\author[e]{H. Cynthia Chiang}
\author[f]{Devin Crichton}
\author[g,h]{Meiling Deng}
\author[i]{Sindhu Gaddam}
\author[e]{Kit Gerodias}
\author[i]{Austin Gumba}
\author[a]{Maile Harris}
\author[i,j]{Kavilan Moodley}

\author[i]{V. Mugundhan}
\author[a]{Laura Newburgh}
\author[k]{Jeffrey Peterson}
\author[e]{Elizabeth Pieters}
\author[a]{Anna R. Polish}
\author[f]{Alexandre Refregier}
\author[i]{Ajith Sampath }
\author[l]{Mario G. Santos}
\author[m]{Onkabetse Sengate}
\author[e]{Jonathan Sievers}
\author[a]{Ema Smith}
\author[a]{Will Tyndall}
\author[i]{Anthony Walters}
\author[n]{Amanda Weltman }
\author[e]{Dallas Wulf}

\affil[a]{Department of Physics, Yale University, New Haven, CT, USA}
\affil[b]{Instrumentation Division, Brookhaven National Laboratory, Upton, NY, USA}
\affil[c]{Department of Computer Science and Electrical Engineering, and Center for Gravitational Waves and Cosmology, West Virginia University, Morgantown, WV, USA}
\affil[d]{Laboratoire d’Astrophysique, Ecole Polytechnique Federale de Lausanne, EPFL, Observatoire de Sauverny, Versoix, Switzerland}
\affil[e]{Department of Physics, McGill University, Montreal, QC, Canada}
\affil[f]{Institute for Particle Physics and Astrophysics, ETH Zurich, Zurich, Switzerland}
\affil[g]{Perimeter Institute for Theoretical Physics, Waterloo, ON, Canada}
\affil[h]{Dominion Radio Astrophysical Observatory, Kaleden, BC, Canada}
\affil[i]{School of Mathematics, Statistics and Computer Science, UKZN, SA}
\affil[j]{Astrophysics Research Centre, UKZN, SA}
\affil[k]{Dept of Physics Carnegie Mellon University, Pittsburgh PA USA}
\affil[l]{Department of Physics and Astronomy, University of Western Cape, Cape Town 7535, South Africa}
\affil[m]{School of Chemistry and Physics, UKZN, SA}
\affil[n]{Department of Mathematics and Applied Mathematics, University of Cape Town, SA}

\authorinfo{Further author information: (Send correspondence to E.R.K.)\\E-mail: emily.kuhn@yale.edu\\}

\begin{document} 
\maketitle

\begin{abstract}
The Hydrogen Intensity and Real-time Analysis eXperiment (HIRAX) aims to improve constraints on the dark energy equation of state through measurements of large-scale structure at high redshift ($0.8<z<2.5$), while serving as a state-of-the-art fast radio burst detector. Bright galactic foregrounds contaminate the 400--800~MHz HIRAX frequency band, so meeting the science goals will require precise instrument characterization. In this paper we describe characterization of the HIRAX antenna, focusing on measurements of the antenna beam and antenna noise temperature.

Beam measurements of the current HIRAX antenna design were performed in an anechoic chamber and compared to simulations. We report measurement techniques and results, which find a broad and symmetric antenna beam for $\nu <$650MHz, and elevated cross-polarization levels and beam asymmetries for $\nu >$700MHz. Noise temperature measurements of the HIRAX feeds were performed in a custom apparatus built at Yale. In this system, identical loads, one cryogenic and the other at room temperature, are used to take a differential  (Y-factor) measurement from which the noise of the system is inferred. Several measurement sets have been conducted using the system, involving CHIME feeds as well as four of the HIRAX active feeds. These measurements give the first noise temperature measurements of the HIRAX feed, revealing a $\sim$60K noise temperature (relative to 30K target) with 40K peak- to-peak frequency-dependent features, and provide the first demonstration of feed repeatability. Both findings inform current and future feed designs.

\end{abstract}


\section{Introduction}
\label{sec:intro}  

HIRAX is a 21\,cm neutral hydrogen intensity mapping experiment and radio interferometer to be deployed at the South African Radio Astronomy Observatory (SARAO) SKA site in the Karoo Desert~\cite{hirax,crichton2021hydrogen}. It operates in the 400--800~MHz frequency range, probing the redshift range $0.8<z<2.5$ to constrain the dark energy equation of state and explore fast radio burst (FRB) and radio transient science. It will eventually consist of 1024 six-meter parabolic dishes \cite{benpaper}, and will map much of the southern sky over the course of four years. Currently, a 10-element prototype array has been deployed at Hartebeesthoek Radio Astronomy Observatory (HartRAO), and a 256-element array is under development at the final HIRAX site.
The demands of 21\,cm observations place tight constraints on the HIRAX instrument, of which the antenna is a crucial part. This paper focuses on two aspects of the HIRAX antenna characterization. It begins in section \ref{sec:antenna} with a discussion of the HIRAX antenna, including design drivers, and simulation work. Section \ref{sec:beams} focuses on beam measurements of several HIRAX antennas in different configurations. 
Section \ref{sec:noisetemp} describes noise temperature assessment of the HIRAX antenna in a facility custom-built at Yale University, following up the work from Kuhn et al.~(2021) \cite{kuhn2021design}.

\section{The HIRAX Antenna}
\label{sec:antenna}
\begin{figure}[h!]
\begin{center}
\includegraphics[width=0.8\textwidth]{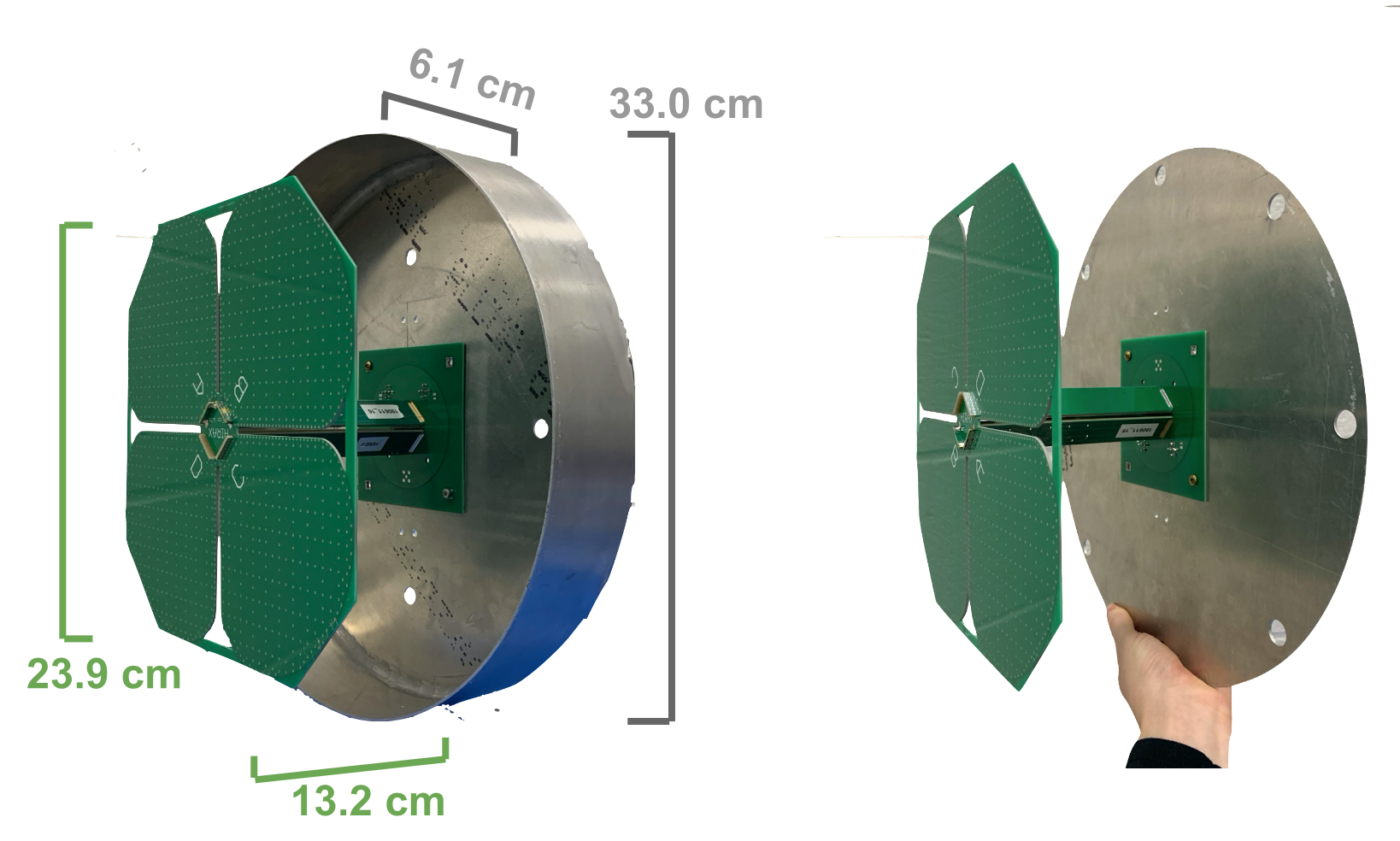}
\end{center}
\caption{The HIRAX antenna, labeled with dimensions, with a can (left) and plate (right) for its ground plane. In operation on the HIRAX dishes, the feed will be mounted on the can structure. For antenna range measurements discussed in section \ref{sec:beams} both configurations are used. Y-factor measurements (section \ref{sec:noisetemp}) are performed only in the plate configuration.}
\label{fig:hiraxfeed}
\end{figure}

Each of the HIRAX dishes will have a dual-polarization antenna mounted at the focus. The HIRAX feed (photo in Figure \ref{fig:hiraxfeed}) is a modified version of the the CHIME antenna\cite{deng14}. It consists of four petals made from FR-4 dielectric (PCB) with an internal copper layer, an active PCB balun, and a support board for coaxial connection.
The petals on the HIRAX feed have rounded edges, providing a distribution of length scales that give a smooth, broadband frequency response. Opposing petal pairs form the arms of a dipole-style antenna, setting the polarization axes and providing a broad beam. A back plane and ring structure (the ``can'') is used to circularize the beam, reduce return loss, and decrease ground spillover and crosstalk to neighboring dishes in the close-packed HIRAX array.

The full antenna is 23.9\,cm in length and width, with a 13.2\,cm long balun, and with the can measuring 33\,cm in diameter and 6.1\,cm in depth. A schematic of the antenna is shown in Figure \ref{fig:hiraxfeed}, with photographs of the antenna mounted to the can as well as to a plate. Both configurations are utilized in characterization tests described in this paper.

The HIRAX design embeds a first-stage differential low-noise amplifier (LNA, Avago MGA--16116) directly into the antenna balun for noise reduction. This design choice has been adopted to keep the total system noise of HIRAX to less than 50\,K, allowing a sensitive measurement of the 0.1mK cosmological 21\,cm signal\cite{shaw2015coaxing}. 

\subsection{Design drivers}
The HIRAX antenna design is constrained by the goal of measuring baryon acoustic oscillations in the redshift range,  $0.8<z<2.5$. It requires the following features: 

\begin{itemize}
    \item Broadband: The antenna should be well-matched in the 400--800~MHz frequency range (redshift range of $0.8<z<2.5$), which probes the window when dark energy is beginning to influence the Universe's dynamics.
    \item Low-noise: A sensitive instrument is needed to measure the 0.1\,mK cosmological 21\,cm signal, which is set against foregrounds that can be hundreds of Kelvin in the relevant radio bands~\cite{chang2008baryon}. For HIRAX, there is a system temperature budget of 50K, of which 30K is contributed by the LNA. 
    \item Low cost: The antenna should be inexpensive to manufacture and build, from the raw materials to the assembly procedure. Minimizing the per-element cost will enable a large N-element array. The cost influences design considerations such as material and weight.
    \item Broad beam: HIRAX utilizes a deep dish design, with F/D = 0.23. Illuminating such a dish requires a broad antenna beam, with low-level sidelobes to reduce spill-over.
    \item Smooth, symmetric beam at each frequency: An effective measurement of the BAO scale will require instrument beam effects to be well understood and then removed in the analysis. A smooth and symmetric beam at each frequency makes measurement and beam deconvolution as straightforward as possible.
    \item Smooth frequency response:
    The foreground mitigation techniques under development for 21\,cm rely on the assumption that foregrounds are smooth in frequency, while the 21\,cm signal is expected to contain spectral structure. To fit foregrounds effectively, instrument properties must be well understood---if the instrument has unexpected frequency structure, smooth foregrounds can no longer be differentiated and removed, as they will contain instrument artifacts\cite{liu2020data}.
\end{itemize}

\subsection{Simulations}

Simulations of the HIRAX feed were run using the commercial electromagnetic modeling software CST Microwave Studio.\footnote{https://www.3ds.com/products-services/simulia/products/cst-studio-suite/} The antenna was modeled as four petals of perfect electrical conductor (PEC), with a PEC balun and can, fed from the center of the cloverleaf, between two opposing petals (see Figure \ref{fig:hirax_feed_cst}). The model does not include the embedded LNA. Feed range measurements in Section \ref{sec:beams} below are compared to simulated beams from this model. The HIRAX feed and dish design and simulations are described in greater detail in Saliwanchik et al. (2021).\cite{benpaper}

\begin{figure}[h!]
\begin{center}
\includegraphics[width=0.6 \textwidth]{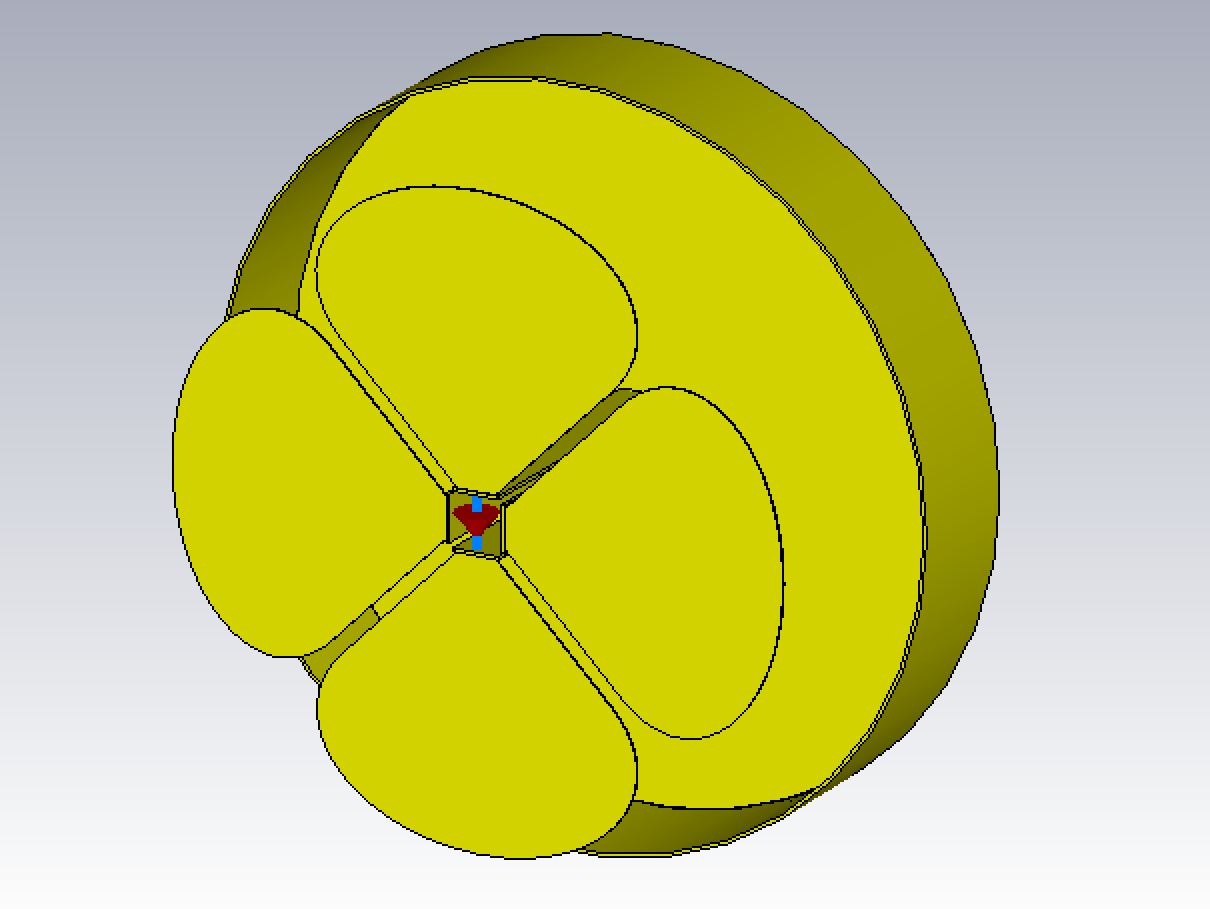}
\end{center}
\caption[]{The HIRAX feed CST model. The simulation feed point is indicated by the blue bar and red arrow, and feeds a single polarization of the dual-polarization antenna.}
\label{fig:hirax_feed_cst} 
\end{figure}

\section{The HIRAX Antenna Beam}
\label{sec:beams}
For antennas that will be used for sensitive cosmological measurements, understanding the beam pattern is a critical characterization step. Artifacts in the antenna such as beam asymmetries or a high cross-polarization component will propagate into the telescope beam and complicate in-field characterization. 
Additionally, measurements of the antenna beam can be used to validate the simulated model of the individual antenna as well as the full telescope. These simulations inform the cosmological forecasts, which feed back into manufacturing requirements. This section describes measurements of the HIRAX antenna beam pattern.

\subsection{Measurement Methodology}

\begin{figure}[h!]
\begin{center}
\includegraphics[width=0.8\textwidth]{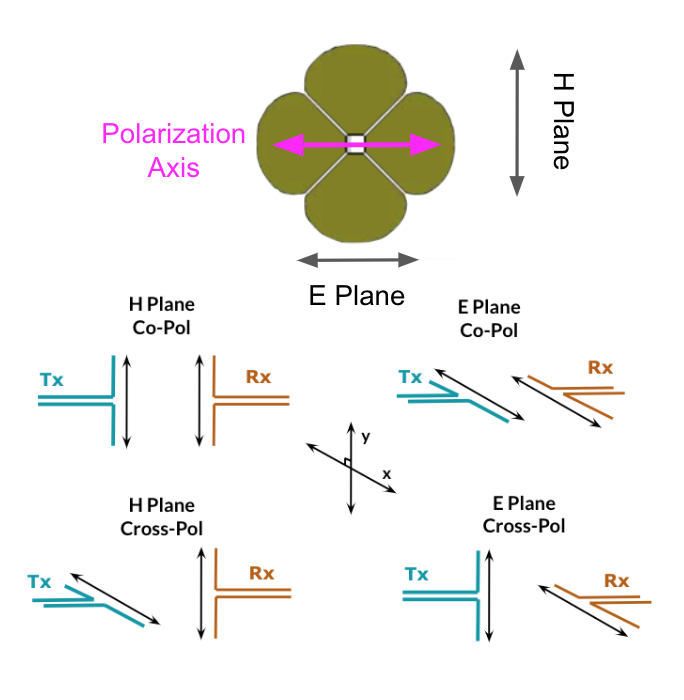}
\end{center}
\caption{Top: the E and H planes for the horizontal dipole element of the HIRAX antenna. Bottom: illustration of antenna orientation for E and H-plane cuts in co and cross-polarization. The antennas are drawn as cartoon dipoles, with the transmitter labeled Tx and receiver labeled Rx. The black arrows indicate polarization direction. During the measurement, the receiver antenna rotates 360 degrees about its y-axis (as pictured in the diagram) while preserving polarization alignment.}
\label{fig:beamcutsdefn}
\end{figure}

For this antenna characterization work, we focus on two orthogonal 1D cuts through the beam. This choice reduces the measurement time per antenna relative to a full 2D map, while still providing the relevant diagnostic information for this design stage. We measure the E and H-plane cuts through the beam, where the E-cut is parallel to the electric field oscillations and the H-cut is parallel to the magnetic field (Figure \ref{fig:beamcutsdefn}). For HIRAX, as for other antennas, the E-plane is parallel to the radiating element---for a dipole this is the plane that contains the two arms---and the H-plane is orthogonal to it. The expected beam pattern for the HIRAX antenna is a broad Gaussian with low-level features at wide angles (called sidelobes) in the E-plane, and a broader beam in the H-plane. These cuts will be measured for both co-polarization, where the transmitting and receiving antenna are aligned in polarization, and cross-polarization, where the transmitting and receiving antenna are orthogonal.

\subsubsection*{Beam measurements in the NSCU Anechoic Chamber}
\begin{figure}[h!]
\begin{center}
\includegraphics[width=\textwidth]{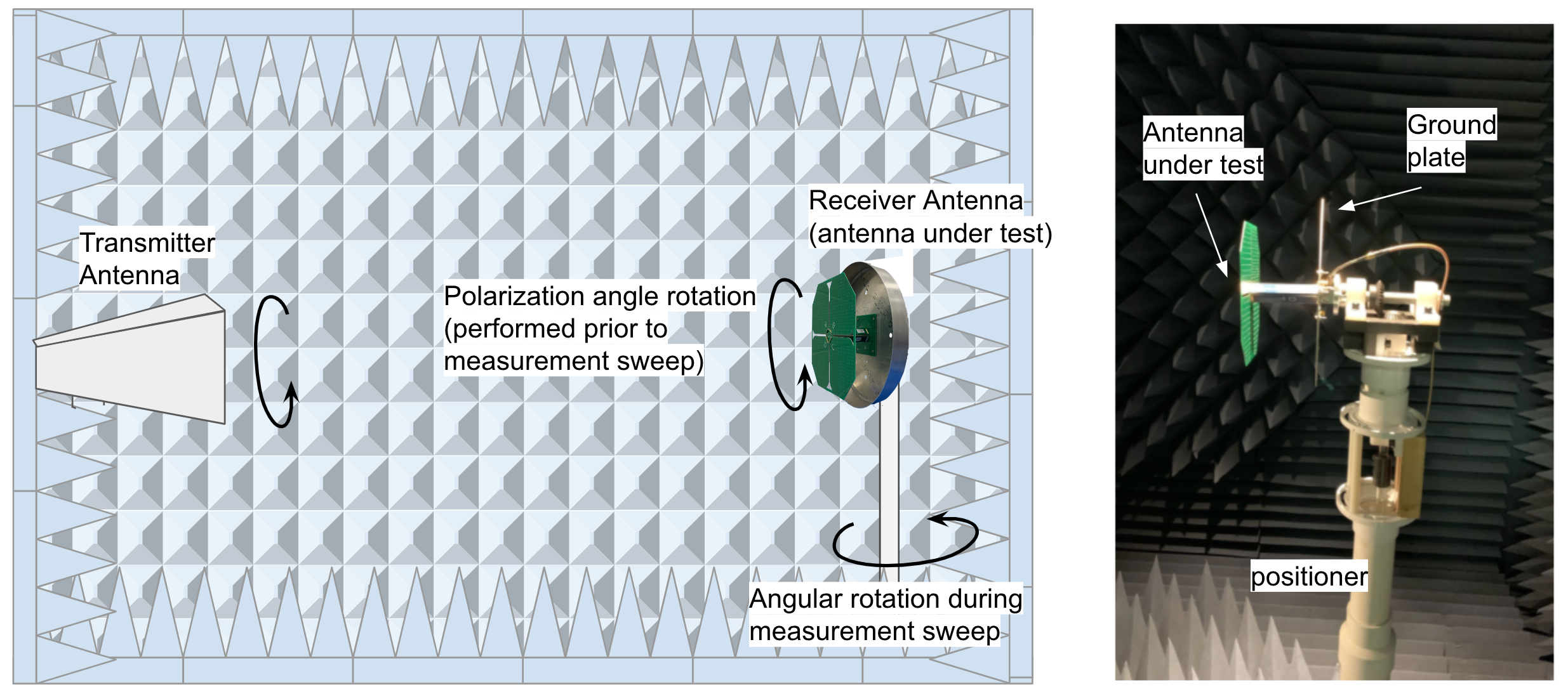}
\end{center}
\caption{Left image: Cartoon depiction of the antenna range measurement set up. Right image: A HIRAX antenna in the anechoic chamber in the North Carolina State University Nanofabrication Facility (NNF), used for antenna beam measurements.}
\label{fig:range}
\end{figure}

The HIRAX antenna beam was measured on an indoor far-field antenna range at North Carolina State University in March 2020, which provided 0.5MHz frequency resolution data from an RF-clean environment. The Nano Fabrication Facility (NNF)\footnote{https://nnf.ncsu.edu/} at North Carolina State University (NCSU) hosts an anechoic chamber for UHF and L-Band antenna testing. The anechoic chamber is 3m by 5m and lined predominantly with 18inch pyramidal RF absorber. In this set up, a transmitting and receiving antenna are separated by a distance of 3.25m. The transmitter pointing direction is fixed, but the transmitter can rotate about its boresight axis. The receiver is mounted on a positioner that enables rotation in azimuth about the boresight axis, as well as rotation about the positioner vertical axis. Prior to a measurement set, a standard gain horn is used to calibrate the transmitter gain, which will be removed from the measurement. For a given measurement set, the two antennas are aligned at boresight along the desired polarization axis, as shown in Figure \ref{fig:range}: for E-plane co-polarization measurements, the radiating elements are aligned and perpendicular to the axis of rotation (parallel to the ground); for H-plane co-polarization measurements, the radiating elements are aligned and parallel to the axis of rotation; for cross-polarization cuts, the radiating elements are anti-aligned.  The positioner then rotates 360 degrees in 3 degree increments, changing the angle between transmitter and receiver while preserving the polarization alignment. A vector network analyzer records a broadband transmission measurement (S21) at each angular position with 0.5MHz frequency resolution. From this series of measurements, a 1D beam pattern is constructed. 

For the HIRAX frequency band (400--800~MHz), the setup uses a standard gain horn optimized for $\sim$700\,MHz as the transmitter. As a result, the transmitter gain curve displays noise at the $\pm$2\,dB level below 450\,MHz. Thus, beam measurements in this paper focus on data measured from 450--800~MHz, and gain measurements specifically focus above 500\,MHz. The cross-polarization transmission from the horn is not characterized directly, though it is bounded at $<-20$\,dB relative to co-polarization through measurements of the CHIME antenna described later in the section.

\subsection{Results}

The measurements and tests were performed as part of the HIRAX antenna beam characterization effort are enumerated in table \ref{tab:measurementlist}. They include: HIRAX antenna beam measurements in two orthogonal 1D cuts for both co and cross-polarization; HIRAX antenna gain measurements; HIRAX antenna beam measurements with a plate as the ground plate, as an investigation of backing structures; and CHIME antenna measurements, for comparison to HIRAX.
The plate and can mounting configurations are shown in Figure \ref{fig:hiraxfeed}.

\begin{table}[h!]
\caption{List of beam cuts performed at NCSU and discussed in this paper. Repeated cuts and rotated measurements are not included.\\}
\begin{centering}
\begin{tabular}{|l|l|l|l|}
\hline
\textbf{Antenna Type} & \textbf{\begin{tabular}[c]{@{}l@{}}Quantity of\\ Feeds Measured\end{tabular}} & \textbf{Measurement Type}                                                                                                           & \textbf{Configuration Notes} \\ \hline
HIRAX                 & 3                                                                           & \begin{tabular}[c]{@{}l@{}}Full beam cuts:\\ (E-plane co-pol, E-plane cross-pol, \\ H-plane co-pol, H-plane cross-pol)\end{tabular} & Mounted to can               \\ \hline
HIRAX                 & 3                                                                           & Gain                                                                                                                                & Mounted to can               \\ \hline
HIRAX                 & 1                                                                           & Full beam cuts                                                                                                                      & Mounted to plate             \\ \hline
CHIME                 & 1                                                                           & Full beam cuts                                                                                                                      & Mounted to plate             \\ \hline
\end{tabular}
\label{tab:measurementlist}
\end{centering}
\end{table}

\subsubsection{HIRAX beam results}
\begin{figure}[h!]
\begin{center}
\includegraphics[width=0.8\textwidth]{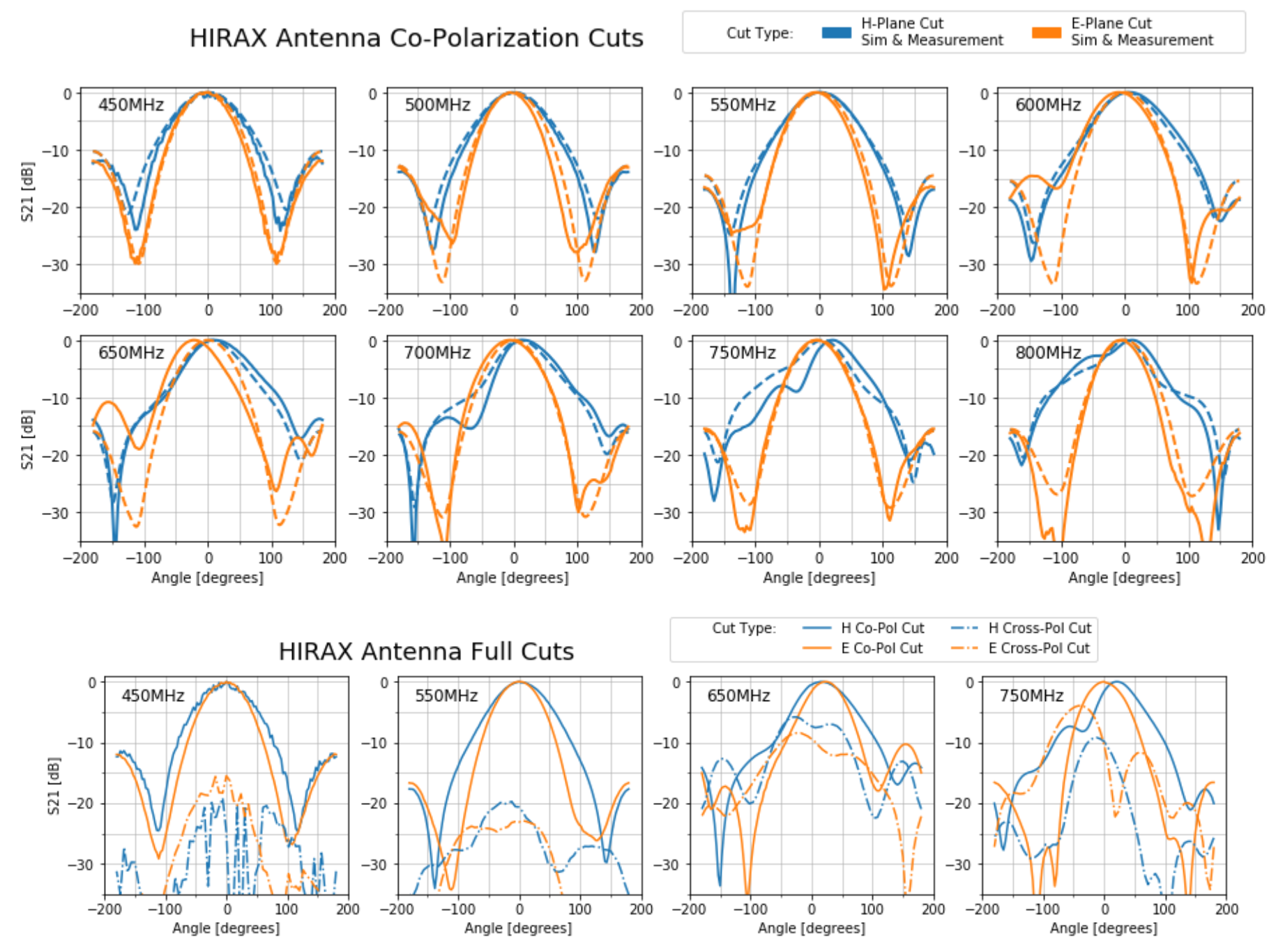}
\end{center}
\caption{Top: Full beam cuts in the E and H-planes for the HIRAX antenna mounted to a can, including measurement (solid line) and simulation (dashed line) results. Measurements are peak-normalized for each frequency. Bottom: Full co and cross-polarization cuts in the E and H planes for the HIRAX feed mounted to a can. The main beam appears to grow asymmetric at higher frequencies, and to develop a high cross-polarization response.}
\label{fig:hiraxcuts}
\end{figure}

Figure \ref{fig:hiraxcuts} shows the HIRAX feed beam for a range of frequencies as measured in the NCSU anechoic chamber, and the corresponding CST simulations. Measurements are peak-normalized for each frequency. In repeated measurements of three separate HIRAX antennas, we recover a beam full-width-half-max in the E-plane that ranges between $88^\circ$--$110^\circ$ with frequency. We observe sidelobe levels between -15\,dB and -10\,dB down from the peak in the E-plane and between -10\,dB and -3\,dB down from the peak in the H-plane, with a dependence on frequency. The measured beams are qualitatively in agreement with simulations. 
The HIRAX feed measurements show frequency dependent beam features (Figure \ref{fig:hiraxcuts}). At lower frequencies ($\nu < 600$\,MHz), both of the E and H-plane cuts appear largely symmetric, and there is roughly 20\,dB of isolation in the main lobe between co- and cross-polarization levels. For $\nu>650$\,MHz, the H-plane centroids begin to drift, by up to 25 degrees, and a secondary peak forms within the main lobe. This centroid location drift can be recovered in drone-based beam measurements of the antenna mounted on a dish, as well as in measurements of source transits over telescope prototypes, which will be discussed in detail in a future paper. The cross-polarization levels are also elevated considerably at high frequencies, and become comparable in magnitude to co-polarization levels within the main beam. 

In simulations we observe beam pointing offsets that correspond closely with the measurements: above 600\,MHz the feed beam centroid varies between 4 and 25 degrees off of boresight, with the maximal deflection at 750\,MHz. We also see significant cross-polarization leakage at high frequencies in simulations. While the boresight cross-polarization response stays 40\,dB below the co-polarization response across the entire band in simulations, this quickly drops off as the angle varies away from boresight. This behavior presents an issue because the main beam of the feed is so broad (FWHM of 70 degrees at 800 MHz). At 50 degrees off from boresight, the main beam has 20\,dB of polarization isolation up to 600\,MHz, but this decreases to 5\,dB of isolation at 700\,MHz, and roughly comparable response in both polarizations from 750-800MHz. The presence of significant cross-polarized response in simulations suggests a solution for this issue.

Possible drivers of the asymmetric beam features are under investigation. These features are likely to be independent of the LNA placement, as simulations do not incorporate the LNA. One likely explanation is that the antenna is fed asymmetrically, building asymmetry into the physical structure, and thus into the beams. 
In simulations, when the feed port is moved to the center of the cloverleaf, both the beam pointing offset and the cross-polarization leakage are completely eliminated. This is strong evidence that the measured beam asymmetries and cross-polarization response are the result of the sligtly asymmetric feed design, and offer a direct path to correcting these issues in the HIRAX feed design. Efforts to design a fully symmetric feeding system for the HIRAX feed are underway, and measurements of updated feed designs will be the subject of future work.

\subsubsection{Measurements with a varied backing structure}

\begin{figure}[h!]
\begin{center}
\includegraphics[width=0.8\textwidth]{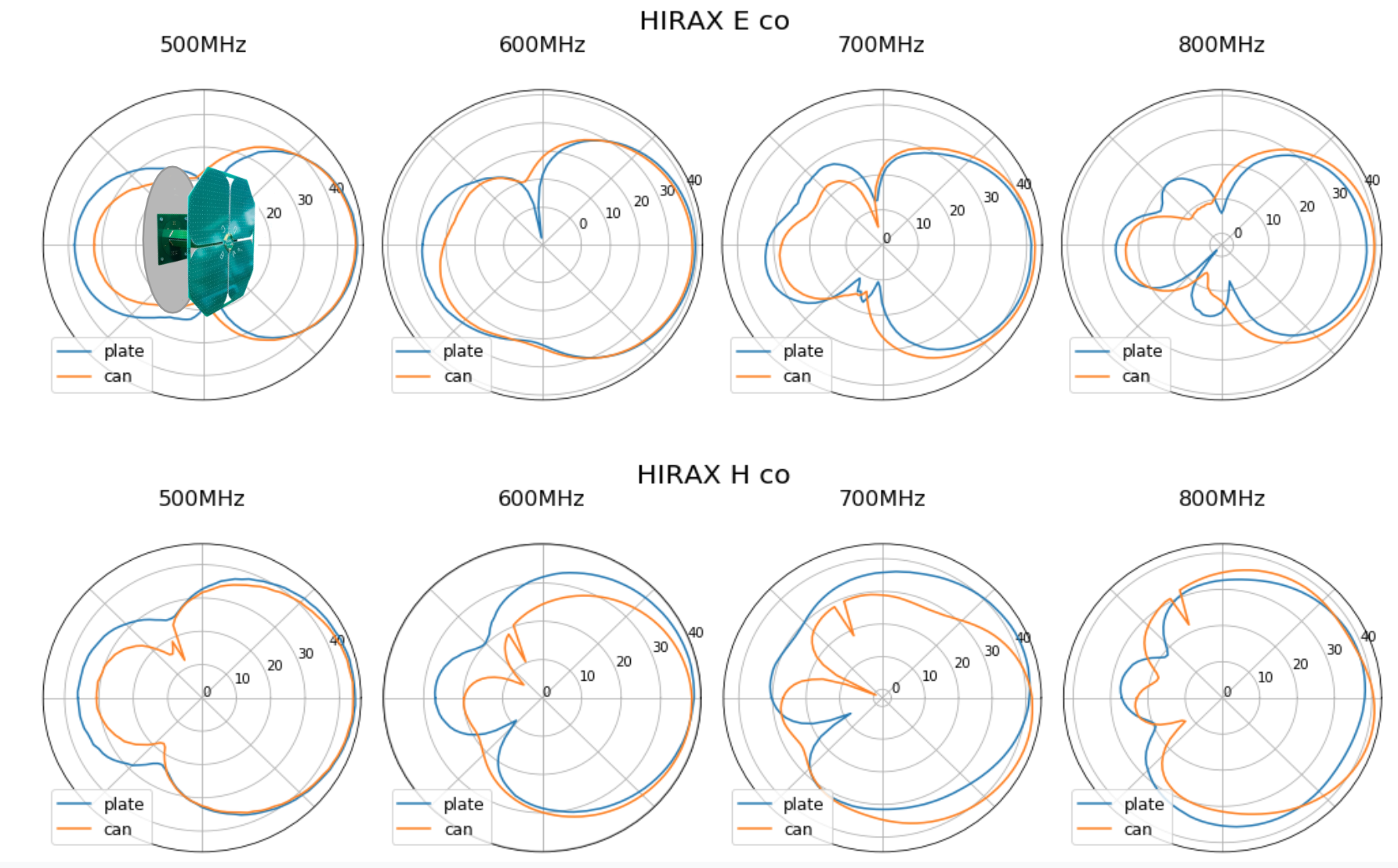}
\end{center}
\caption{Cuts in the E- and H-planes for the HIRAX antenna backed by either a plate (blue) or a can (orange). Photos of the configurations can be found in Figure \ref{fig:hiraxfeed}.} 
\label{fig:platecan}
\end{figure}

The HIRAX feed design utilizes a can structure as a ground plane to circularize the beam and improve its directionality. We verify the can's impact, and perform checks on simulations by performing beam measurements of the HIRAX feed backed by both a can and a plate, and comparing the two. This is important as the HIRAX feed is designed to be mounted on some ground plate structure. Figure \ref{fig:platecan} shows that the introduction of the HIRAX feed can instead of a plate of comparable dimensions can shrink the antenna backlobe by as much as 6\,dB and improve its directionality, as designed. The HIRAX feed and plate are displayed over the 500MHz E-plane co-polarization measurement to illustrate the antenna orientation for the reader.

\subsubsection{HIRAX gain and antenna repeatability}
\begin{figure}[h!]
\begin{center}
\includegraphics[width=\textwidth]{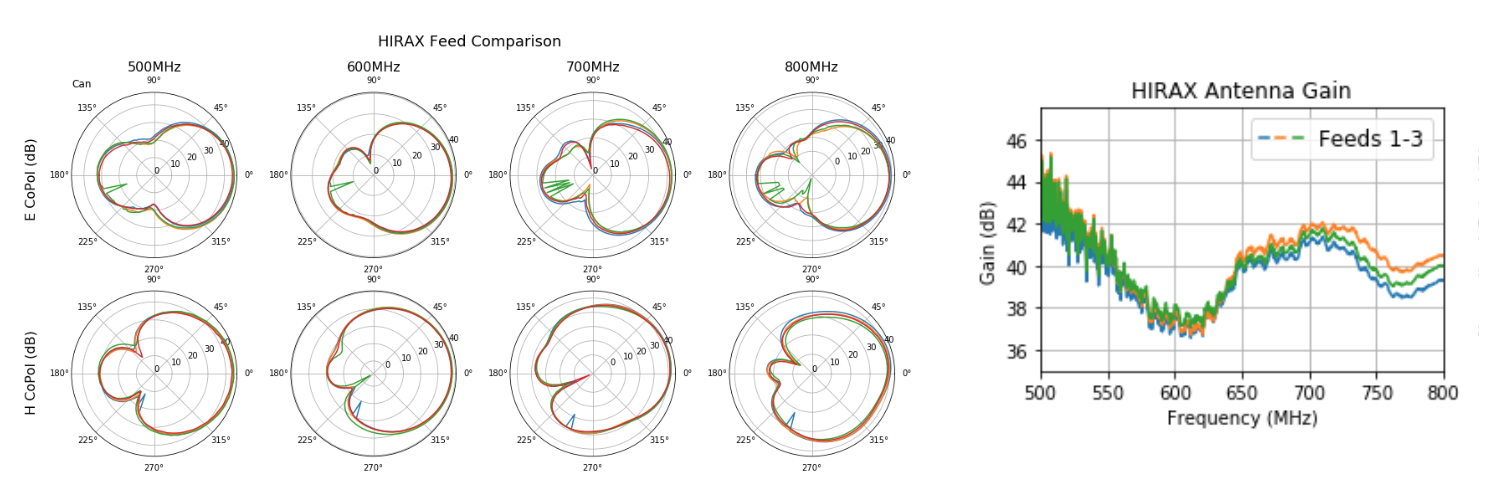}
\end{center}
\caption{Co-polar beam cuts (left) and gain profiles (right) for three HIRAX antennas with integrated first stage LNAs. The antenna beams are similar at the sub-dB level.}
\label{fig:feedcompare}
\end{figure}

The HIRAX feed has an average gain of $\sim40$\,dB that varies by up to 3\,dB in frequency, shown in Figure \ref{fig:feedcompare}. The gain was assessed during beam measurement sweeps, as the broadband transmission when the antenna under test was oriented at 0 degrees relative to the transmitter. The measured gain is a combination of the antenna gain resulting from feed geometry (expected 6-8dB) and the gain of the embedded LNA (expected 35-40db). The measured gain profiles are calibrated to remove the transmitter beam profile and cable effects, as well as account for free space path loss.

Antenna beam measurements were repeated for three different HIRAX feeds (with a can ground plane) to assess antenna manufacturing repeatability, and the results are shown in Figure \ref{fig:feedcompare}. The measured gain was found to have up to 1\,dB variations, with sub-dB variations across $\nu<700$MHz. This is within measurement error, as determined by repeated gain measurements of the same antenna antenna displaying comparable fluctuations at high frequencies (possibly due to alignment or calibration error).
The beam patterns of the three HIRAX antennas are qualitatively similar, and vary within 2\% in full-width-half-maximum as obtained from Gaussian fits.
The implications for antenna repeatability are important: the HIRAX instrument requirements are generally more concerned with precision than accuracy, as the array operation relies on element redundancy\cite{crichton2021hydrogen}.

\subsubsection{Comparison to CHIME feed measurements}
\label{sec:chimedescription}
The CHIME antenna design was the basis for the HIRAX antenna\cite{deng14,chime2}, and both share a similar but not identical design. 
The CHIME and HIRAX antennas are both broadband and optimized for performance in 400--800~MHz frequency band. In contrast to the HIRAX antenna, the CHIME antenna is made from teflon, a lower loss (but higher cost) material than FR4, which allows the first stage amplifier to be placed after the antenna structure while maintaining an acceptable noise temperature. Polarization in the CHIME antenna is summed along an axis rotated 45 degrees with respect to the HIRAX antenna polarization.

\begin{figure}[h!]
\begin{center}
\includegraphics[width=\textwidth]{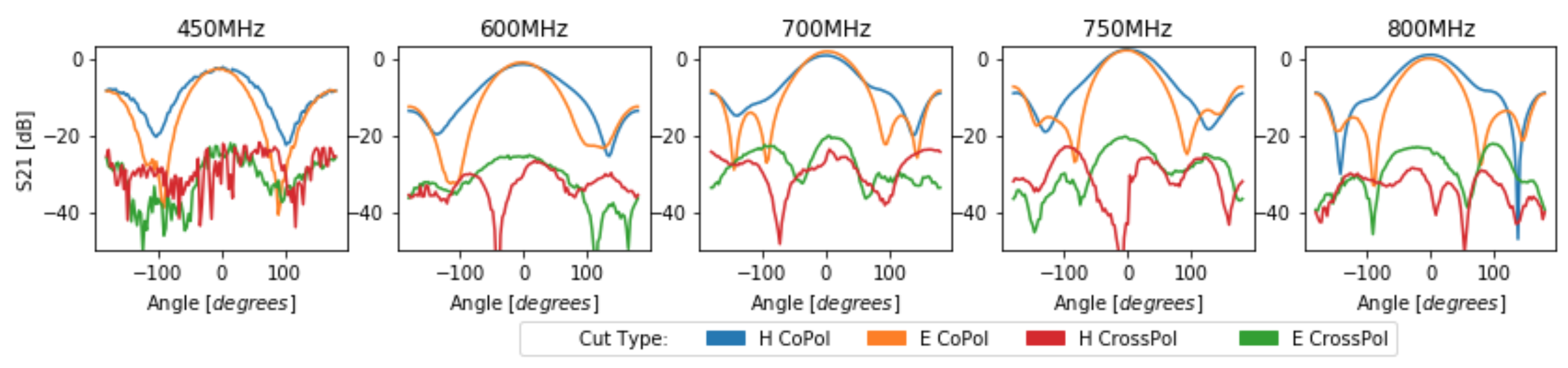}
\end{center}
\caption{E and H-plane cuts for the CHIME antenna, including both co and cross-polarization measurements. This antenna is similar in design to the HIRAX antenna, but displays improved symmetry in the co-polarization beam, and lower levels of cross-polarization.}
\label{fig:chimecuts}
\end{figure}

Beam measurements of the CHIME antenna with a circular ground plate were performed, and results were compared qualitatively with HIRAX antenna measurements in the same mounting configuration.
The full cuts of the CHIME feed (Figure \ref{fig:chimecuts}) show a broad, smooth, gaussian beam  that is generally symmetric, with low-level cross-polarization. 

In measuring full cuts of the CHIME feed, there are revealing contrasts with the HIRAX feed performance. While asymmetries do arise at high frequencies in the CHIME antenna beam, they are not nearly as stark as in the HIRAX antenna, and the centroid drift is much less substantial (fewer than 5 degrees for CHIME, while up to 25 degrees for HIRAX). The level of cross-polarization for the CHIME antenna remains $<-20$\,dB relative to the co-polarization component in the main beam; at frequencies $\nu > 700$MHz, the HIRAX antenna displays cross-polarization levels that can meet or exceed the co-polarization level for parts of the main beam.

These observations indicate that the high level of cross-polarization observed in the HIRAX antenna beam is associated with the differences between the CHIME and HIRAX feed designs. This comparison suggests that with appropriate design modifications HIRAX may be able to accomplish an improved antenna performance. 

\subsection{Systematics and Uncertainties}
Several repeated measurements were performed in order to capture measurement fluctuations and bound statistical uncertainty. We find $<1$\,dB variation in the main beam between 3 sets of repeated measurements. 

Measurements were performed to assess the impact from reflections in the anechoic chamber on asymmetries present in the data. Range reflections can be accounted for by exploiting a simple technique in which beam measurements are repeated with the AUT rotated by $180^{\circ}$ in polarization angle, such that the polarization axes remain parallel between the transmitter and receiver but the antenna orientation is now flipped upside-down relative to its starting position. 
If the asymmetries in the beam measurements are due to the antenna, they will rotate by 180 degrees (amounting to a reflection about $x=0$) upon rotation of the antenna by 180 degrees. If the asymmetries are due to reflections in the anechoic chamber, they will not rotate upon rotation of the feed.
This assessment is particularly important in the NCSU setup, as the 18in absorber used in the anechoic chamber is rated for GHz frequencies, and its efficiency may be reduced in the HIRAX band. 

\begin{figure}[h!]
\begin{center}
\includegraphics[width=0.8\textwidth]{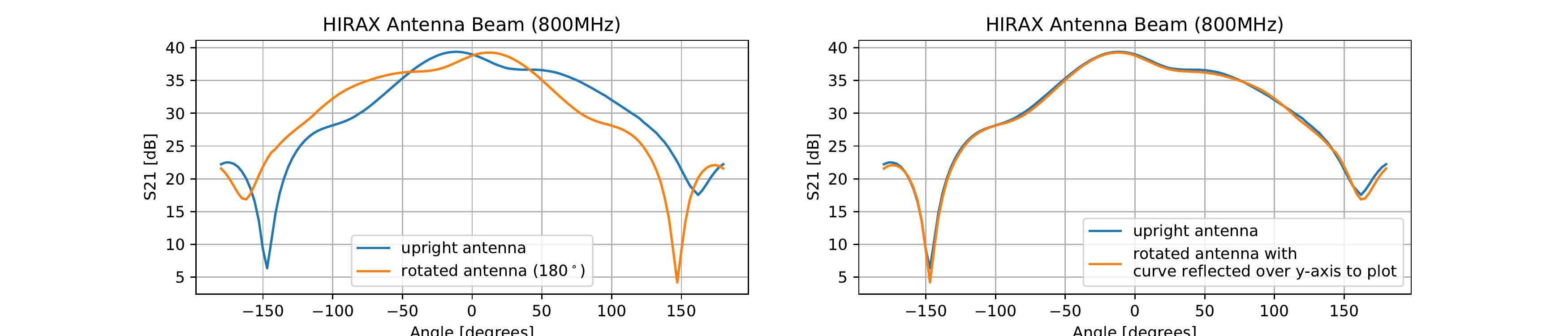}
\end{center}
\caption{H-Plane co=polarization beam measurements of the HIRAX antenna at 800MHz, repeated for the antenna flipped upside-down (rotated 180$^\circ$ about the axis between the receiver and transmitter antennas). Measurements are taken in both configurations to verify that beam asymmetries are true features of the antenna and not artifacts of the antenna range. To do this, we complete a beam cut, then repeat it with the feed rotated by 180 degrees. The two cuts are plotted as the blue and orange curves (left plot). If the orange curve is mathematically flipped about $x = 0^\circ$, we find a near-identical signal to the blue curve (right plot), which means that the asymmetries are due to the feed performance and are not an artifact of reflections in the environment.}
\label{fig:flipped}
\end{figure}

Applying the described methodology to the HIRAX antenna beam, which displays asymmetric beam features at frequencies above 600MHz, we find anechoic chamber reflections to contribute sub-dB level uncertainty at all frequencies in the measurement range, which is within the statistical uncertainty bound found from repeated measurements. Figure \ref{fig:flipped} shows an example of this result at 800MHz, with the blue and orange curves showing co-polarization measurements made with the feed oriented at 0 degrees (upright) and 180 degrees (rotated) relative to the polarization axis. For each curve there is a double-lobed main beam (lobes at -14, 50 degrees for the upright antenna and -50, 14 degrees for the rotated antenna) with the lobes differing in amplitude by 2.5dB. The locations of the first nulls are shifted by $\pm\sim$8 degrees from symmetry for each curve and by 15 degrees between the two curves (nulls are located at -147, 162 degrees for the upright antenna, and at -162, 147 degrees for the rotated antenna). When the rotated antenna measurement is reflected about $x = 0$, both measurement curves align to dB-level, validating that the beam is truly asymmetric and measurements are not hindered by the test environment. Lower frequencies ($\nu < 600$MHz) where the absorber is expected to perform worse do not show degradation in the main beam agreement, indicating that the absorber was still effective at the low frequencies presented in this paper. This test does not apply to frequencies below 500MHz, where the beam is sufficiently symmetric that a rotation about $x=0$ degrees does not change the curve.

In this dataset, we focus on understanding the HIRAX antenna beam shape, relative levels of features in the beam (e.~g.~ sidelobe level to the main beam, between feeds, etc), relative amplitude of co- and cross-polarization beams, and relative gain between different antennas. 
An absolute calibration of the telescope will be performed in-situ, so we do not focus on systematic effects that add standard offsets in this analysis. We must still quantify systematic errors that could modify stages of the measurement differently to introduce beam structure. This would be dominated by reflections, which are described above, and found to contribute within the statistical uncertainty bound. We find the reported beam measurements to be reliable to 1\,dB.

\section{Antenna Noise Temperature}
\label{sec:noisetemp}
The HIRAX system noise has a target of $T_\text{sys} = 50 \pm 5$K, with $<30$K contributed by the antenna. 
The HIRAX antenna design is optimized to lower noise by locating the first stage amplification in the antenna structure. As a result, the LNA noise temperature cannot be directly measured with a noise figure meter, and thus we must use a free-space measurement setup for full assessment of the HIRAX antenna temperature. 

To measure the feed noise temperature, we  take a Y-factor measurement\cite{pozar}, which compares relative output power ($Y={P_{\rm hot}}/{P_{\rm cold}}$) at two temperatures (${T_{\rm hot}, T_{\rm cold}}$) as measured by some test device to determine the device noise temperature $T_{\rm noise}$:

\begin{equation}
Y=\frac{P_{\rm hot}}{P_{\rm cold}}
\end{equation}

\begin{equation}
T_{\rm noise}=\frac{T_{\rm hot}-YT_{\rm cold}}{Y-1}
\end{equation}
with $Y$ referred to as the ``Y Factor.'' 

This type of measurement is commonly taken using the sky as a cold load, which is difficult in the RFI rich environments around universities. Additionally, the HIRAX antenna beam is too broad to sufficiently understand the temperature of each load in such a measurement. 
We have thus built a mechanism at Yale University for convenient antenna noise temperature measurements, that will provide feedback for current and future feed iterations. This system has been designed, fabricated, and tested for agreement with simulations (described in Kuhn et al. (2021)\cite{kuhn2021design}), and has been used to produce the first noise temperature measurements of the HIRAX antenna, described in this section. 

\subsection{System design}  
The system design is described in detail in Kuhn et al.~\cite{kuhn2021design}. Here, we summarize and note updates from that publication. 

\begin{figure}[h!]
\begin{center}
\includegraphics[width=0.8\textwidth]{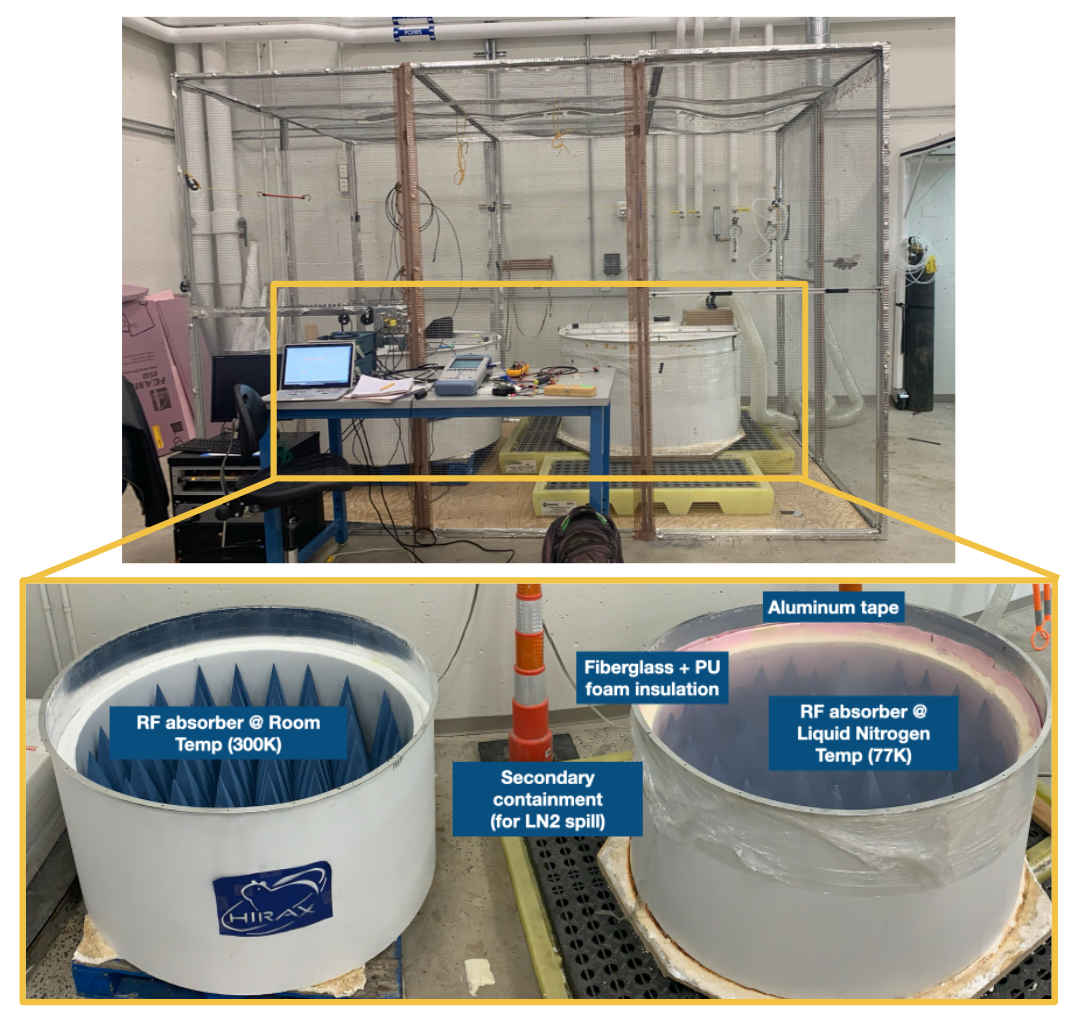}
\end{center}
\caption{Top: The Y-factor measurement system, including a custom Faraday cage enclosing the hot/cold loads. The cage was constructed for RFI mitigation, and demonstrations show it provides 20\,dB of attenuation for unwanted signals. The cage additionally serves as an extra safety barrier between the nitrogen bath and spectators in the lab, and the structure provides a convenient way to support cables and relieve cable strain. Bottom: Labeled photograph of the hot/cold loads for noise temperature tests, taken during an LN2 fill. The loads are covered and sealed with a steel lid for measurements.}
\label{fig:yfacsystem}
\end{figure}

The antenna temperature measurement system consists of two steel cylinders, 51 inches in diameter and 26 inches in height. They are lined with insulating foam, and the bottoms covered with 18 inch pyramidal RF absorber (Figure \ref{fig:yfacsystem}). During measurements, one of the cylinders remains at room temperature (295K) and the other is filled with liquid nitrogen (77K), submersing the RF absorber and providing two black bodies at different temperatures. An insulating cover sits between the liquid nitrogen surface and the antenna, to prevent it from cooling. The laboratory ventilation system is additionally used to draw the cold nitrogen gas (from the liquid boil off) away from the antenna, which maintains the antenna at physical temperatures above 285K such that the cooling has no discernible impact on the measurement. 

A faraday cage has been built to enclose the Y-factor system for RFI mitigation purposes, and is an upgrade from our 2021 publication\cite{kuhn2021design}. The cage is 12\,ft$\times$8\,ft$\times$8\,ft, with a frame made of aluminum T-slot material\footnote{https://www.mcmaster.com/t-slotted-framing/}. Panels of 0.5 inch steel mesh are bolted onto the frame and then electrically connected by layers of aluminum tape with conductive adhesive. In measurements of transmission between an antenna in the cage and an antenna outside of it, the cage was determined to provide 20\,dB isolation from stray RF signals, allowing for a cleaner noise temperature measurement. The measurement system is shown in Figure \ref{fig:yfacsystem}.  

\subsubsection{Y-factor measurement procedure}
The antenna under test is mounted to a 12\,in\,$\times$\,12\,in ground plate that slots into the cavity lid. This mounting plate is bolted onto the ``hot'' cylinder lid, such that the antenna is enclosed in the system and facing the RF absorber. The antenna output is connected to a spectrum analyzer and data-taking is initiated, integrating with 3\,MHz bandwidth for 1s total integration time and saving a file every 10 seconds for $\sim5$ minutes. After the  $\sim5$ minutes of ``hot'' measurements, the antenna and plate are moved over to the cold cylinder and bolted into its lid, where data is recorded for $\sim 5$ minutes of ``cold'' measurements. This transition between hot/cold measurements is repeated three times to assess the impacts of antenna placement, environmental fluctuations, and related factors. For all data sets, the hot measurements to are separately averaged and the cold measurements are separately averaged before computing a noise temperature and associated statistical error, and the there are dozens of hot/cold measurements included in the statistics.

\subsection{Hot/cold verification}
Initial verification measurements were performed at room temperature to quantify sources of systematic error and assess whether the cylinders are similar to within specifications. These tests are described in Kuhn et al.~(2021)\cite{kuhn2021design}, and project liquid nitrogen surface reflections and RF differences between the cavities to contribute 3.5-8K systematic offsets that can be removed. We extend this work to include verification measurements with the full cryogenic system, discussed in this paper. 

For verification measurements of the cryogenic system, we measure the noise temperature of various amplifiers attached to the CHIME antenna, described in section \ref{sec:chimedescription}. The noise temperatures of the commercial amplifiers are independently measured with two commercial noise figure meters, and the gains are measured with both a noise figure meter and VNA. The CHIME antenna is low-loss and expected to contribute 8-10\,K to the system noise based on CST simulations.
We can compare noise temperature measurements of the amplifiers in the Yale system to the known amplifier noise temperature (as determined with a noise figure meter) to assess systematic errors.

\subsubsection{Methodology (Verification Tests)}
We measured the noise temperature of 3 commercial amplifiers: Mini Circuits ZX60-112LN+ (`112' hereafter), ZX60-P103LN+ (`P103' hereafter), and ZX60-P105LN+ (`P105' hereafter). The noise temperatures of these amplifiers were characterized with a commercial noise figure analyzers (NFA). 

The NFA used for the measurement was an Agilent N8975 Noise Figure Analyzer. We connected an HP346B noise source, entered the ENR table displayed on the noise source, and followed the standard calibration procedure. The amplifier under test was then attached and NFA data was taken across 400--800 \,MHz assuming an ambient temperature of 290\,K. The noise temperature measurements contained a sinusoidal ripple pattern across the observed frequency range which indicated a mild impedance mis-match; we found that installing 6\,dB attenuators on the input of the amplifier reduced the reflections, and we accounted for the additional noise temperature increase by subtracting the attenuation values from the ENR table and re-doing the calibration procedure.  The noise temperatures from these measurements are given in Table~\ref{t:NTres}. Statistical errors in the data sets are the standard deviation over the repeated measurements and are typically 0.3-1.2\,K. The systematic errors are the differences in noise temperature measured with attenuation settings and set up configurations; they are typically 0.6-2\,K, though can range up to 5K for the 112 amplifier in a subset of the band. These values are within the NF analyzer reported uncertainty of 0.05\,dB in NF (3--5\,K for this amplifier set, depending on the amplifier measured) when used with our specific calibration source. 

\begin{table}[ht]
\centering
\begin{tabular}{|l|l|l|l|}
\hline
\textbf{Amplifier Part \#} & \textbf{Gain (dB)} & \textbf{Noise Figure (dB)} & \textbf{Noise Temp (K)} \\ \hline
ZX60-112LN+     & 26.40 - 26.68      & 1.03 - 1.11                & 78- 84 $\pm$ 2K                \\ \hline
ZX60-P103LN+                   & 16.84 - 21.44      & 0.49 - 0.68                & 34 - 49 $\pm$ 2K               \\ \hline
ZX60-P105LN+                   & 14.8-14.6              & 1.81 - 1.96                & 150-165 $\pm$ 2K               \\ \hline
\end{tabular}\\
\vspace{4pt}
\caption{\label{t:NTres}Noise temperature and gain in 400--800~MHz for the three commercial minicircuits amplifiers used for cryogenic system verification. These values were measured using a noise figure meter and vector network analyzer.}
\end{table}

\begin{figure}[h!]
\begin{center}
\includegraphics[width=\textwidth]{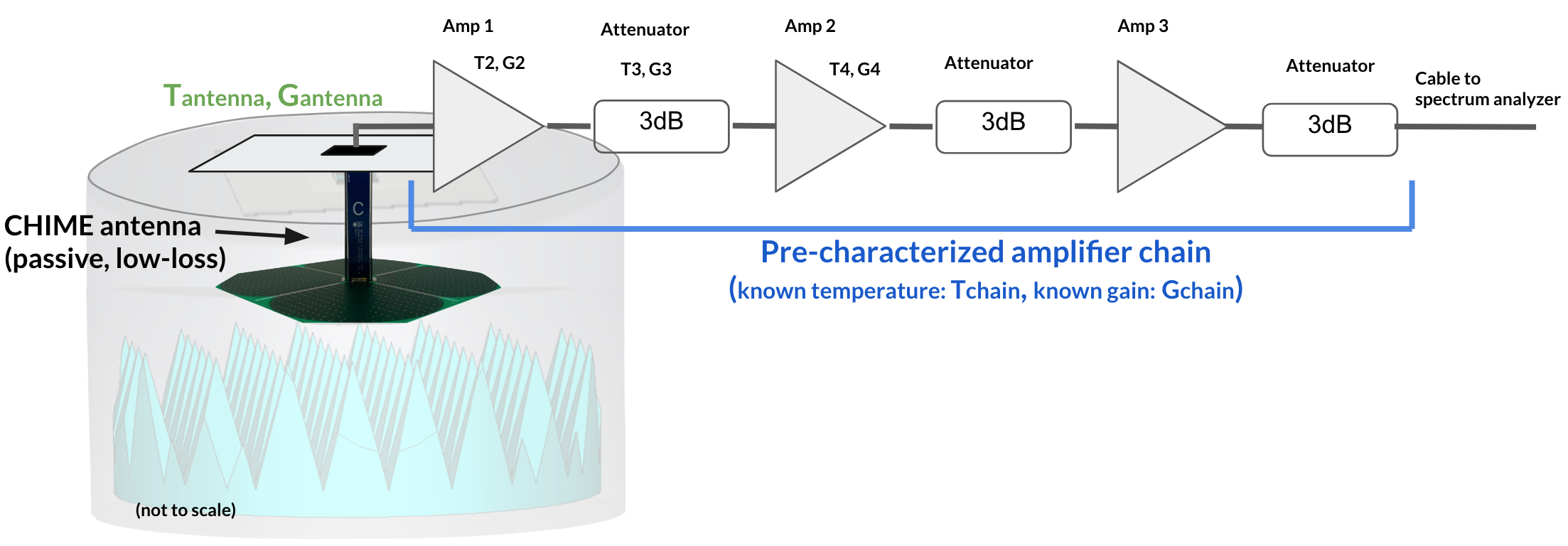}
\end{center}
\caption{Measurement set up used for the  for the cryogenic verification measurements. In this set up, we measure the noise temperature of a passive feed (with temperature $T_\text{antenna}$ and gain $G_\text{antenna}$) connected to a chain of previously characterized amplifiers (with temperature $T_\text{chain}$ and gain $G_\text{chain}$). The amplifier chain includes attenuators between active components and long cables to limit reflections. From this measurement, the feed loss can be recovered and compared to an expected value.}
\label{fig:hotcoldsystem}
\end{figure}

Using the characterized amplifiers, we can perform measurements of the CHIME feed (plus amplifiers) in the full cryogenic system and compare that result to expectations from simulations. The procedure is as follows. 

The noise temperature $T_{1}$ of a passive element at ambient temperature $T_\text{ref}$ with loss $L$ can be expressed as,
\begin{equation}
    T_{1} = T_\text{ref}(L-1),
\label{eq:tempandloss}
\end{equation}
with $T_\text{ref} = 290$K. Loss ($L$) and gain ($G$) are related by,
\begin{equation}
    G = \frac{1}{L}.
\label{eq:gainandloss}
\end{equation}
The equation for noise temperature of cascading devices can therefore be re-written in the following way for our measurement system, which includes a passive antenna (noise temperature $=T_\text{antenna}$, gain $ = G_\text{antenna}$) as the first element, and a chain of amplifiers and attenuators (noise temperature $ = T_\text{chain}$) as subsequent elements (Figure \ref{fig:hotcoldsystem}): 
\begin{equation}
\begin{split}
T_\text{noise} & = T_1+\frac{T_2}{G_1}+\frac{T_3}{G_1G_2}+ ... \\
T_\text{noise} & = T_\text{antenna}+\frac{T_\text{amp1}}{G_\text{antenna}}+\frac{T_\text{amp2}}{G_\text{antenna}G_\text{amp1}}+ ... \\
& = T_\text{antenna} + \frac{1}{G_\text{antenna}}\times \left(T_\text{amp1}+\frac{T_\text{amp2}}{G_\text{amp1}}+ ...\right)\\
& = T_\text{antenna} + \frac{T_\text{antenna}+290K}{290K}\times T_\text{chain}\\.
\end{split}
\label{eq:cascadent}
\end{equation}
We combine equations \ref{eq:tempandloss} and \ref{eq:gainandloss} to write $G_\text{antenna}$ in terms of $T_\text{antenna}$, and assume ohmic losses as the primary source of loss.

\subsubsection{Results from Hot/Cold Verification Tests}
We can solve for the effective CHIME antenna noise temperature $T_\text{antenna}$ from a Y-factor measurement of the CHIME feed and known amplifier chain ($T_\text{noise}$), a noise figure meter measurement of the amplifier chain alone ($T_\text{chain}$), and an assumed ambient temperature ($T_\text{ref}$), following methods above (Figure \ref{fig:hotcoldsystem}), 
\begin{equation}
    T_\text{antenna} = T_\text{ref}\left(\frac{T_\text{noise}+290}{T_\text{chain}+290}-1\right).
\end{equation}

We solve for the CHIME antenna temperature (and measurement system contributions) from measurements with three different amplifier chains, finding that it ranges from $\sim$5-15K with common frequency dependent features and is recoverable to within 3K for each chain. This noise temperature would be from a combination of factors---material loss in the feed itself, impedance mismatches in the measurement chain, and reflections in the noise temperature system---and should not be interpreted as a direct measurement of the CHIME feed loss. CST simulations of the CHIME antenna (with SMA connector) predict a 8-10K antenna temperature using radiation efficiency as a proxy for antenna loss, which is consistent with the $\sim$5-15K measured temperature plus $\pm$5K error. Comparing the simulated contribution from the CHIME antenna with the $T_{antenna}$ inferred from Yale system measurements bounds the systematic error contributed by the measurement set up to be less than 5\,K.

\begin{figure}[h!]
\begin{center}
\includegraphics[width=0.8\textwidth]{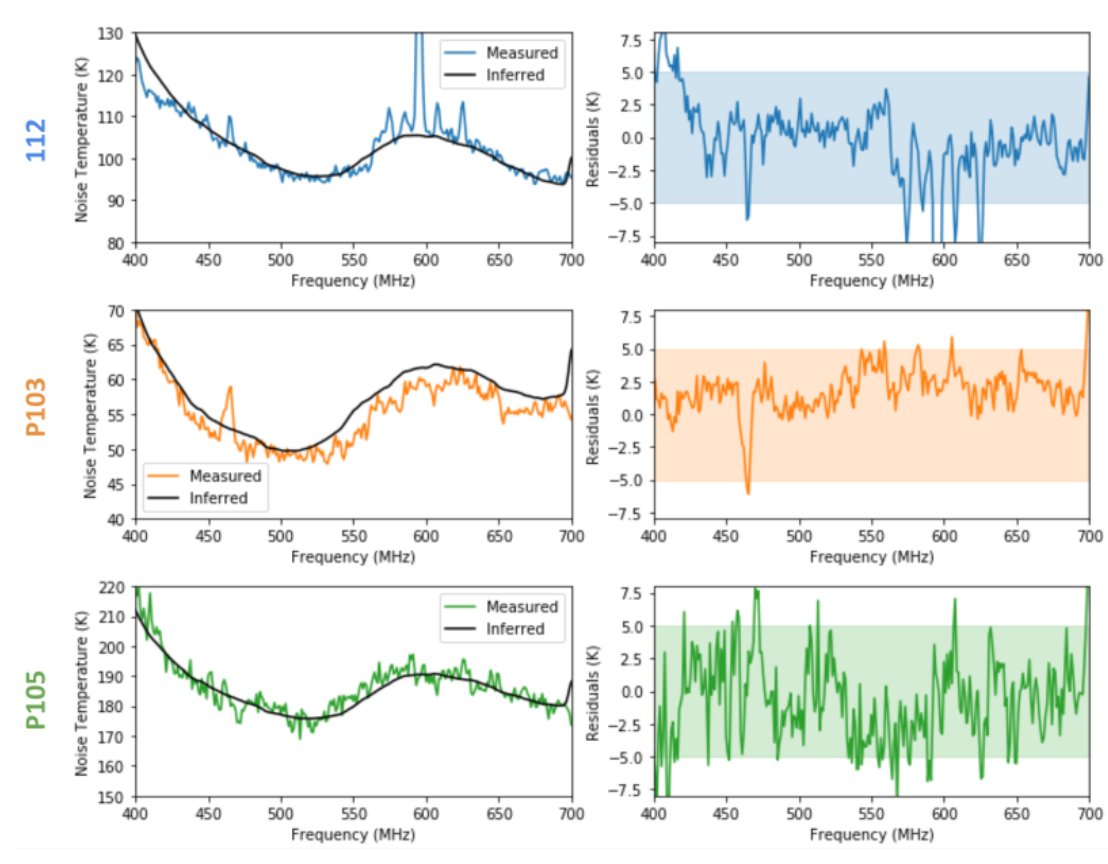}
\end{center}
\caption{Y-factor measurements and inferred results for CHIME antenna plus several amplifier chains. The inferred results were determined from a noise temperature evaluation of the CHIME feed alone, in combination with noise figure measurements of the amplifier chain components (shown in Figure \ref{fig:hotcoldsystem}). The left hand plots show the noise temperature comparisons for each of the three amplifier chains (first-stages ZX60-112LN+, ZX60-P103LN+, and ZX60-P105LN+), and the right hand plots show the difference between measured and expected along with a $\pm$5K shaded region. The standard deviation of the residuals gives 1-sigma error bounds below 3.3K for all three amplifiers.}
\label{fig:chimeresults}
\end{figure}

Figure \ref{fig:chimeresults} shows the noise temperature of the CHIME antenna plus different amplifier chains, with the legend indexed by the first LNA in each chain. The measured noise temperature is compared with an inferred noise temperature (black line), constructed from equation \ref{eq:cascadent} with: $T_\text{antenna}$ as a spec on CHIME antenna temperature, obtained by averaging the three $T_\text{antenna}$ results from noise temperature measurements with different amplifier chain measurements; and $T_\text{chain}$ from existing knowledge of the amplifier chain noise, obtained from noise figure meter measurements of individual amplifiers. With the CHIME feed measurements, we are able to recover the measured noise temperature to within $\pm$5K of the expected values for all three measured amplifier chains (Figure \ref{fig:chimeresults}). This shows that after applying a correction for the feed and measurement losses, we can recover noise temperatures within our 5\,K performance criteria. 
The standard deviation of the residuals gives 1-sigma error bounds of 2.3K for the 112 amplifier measurement (excluding the high-RFI region 550--650~MHz), 1.6K for the P103 amplifier measurement (no regions excluded), and 3.3K for the P105 amplifier measurement (no regions excluded. These error bounds are comparable in performance to the commercial noise figure meter (2K error). 
We also note that the computed error is a common offset between the measurements and not fractional, with the fractional uncertainty ranging from 3-9\% between the three amplifier chains.

A noise temperature measurement was performed for the 112-amplifier chain connected to a short cable and SMA connector and terminated with a 50$\Omega$ terminator. The thermal noise of this load, amplified by the amplifier chain, was measured at room temperature (290K) and with the connector dunked in a container of liquid nitrogen (77K). 
The connector noise temperature is inferred to be 10K, with frequency structure of amplitude 3K. The 10K noise temperature corresponds to a loss of roughly 0.15\,dB, which was verified with our VNA `cable loss' setting. This value is 0.1\,dB higher than published data sheet values for a standard connector\footnote{https://www.pasternack.com/images/ProductPDF/PE9505.pdf}, and we expect with the inclusion of cable loss and terminator noise the value would agree.  
The CHIME loss is expected to be in this $10K$ regime, as it is made of low-loss material and includes an SMA connector, and we see it averages to the same general value but with higher amplitude features.
This result gives further confidence in the overall measurement levels.

\subsection{Noise Temperature Results}
We have taken Y-factor measurement data for four different HIRAX feeds over the course of five separate measurement trials, from December 2019 through December 2021.

\begin{figure}[h!]
\begin{center}
\includegraphics[width=0.8\textwidth]{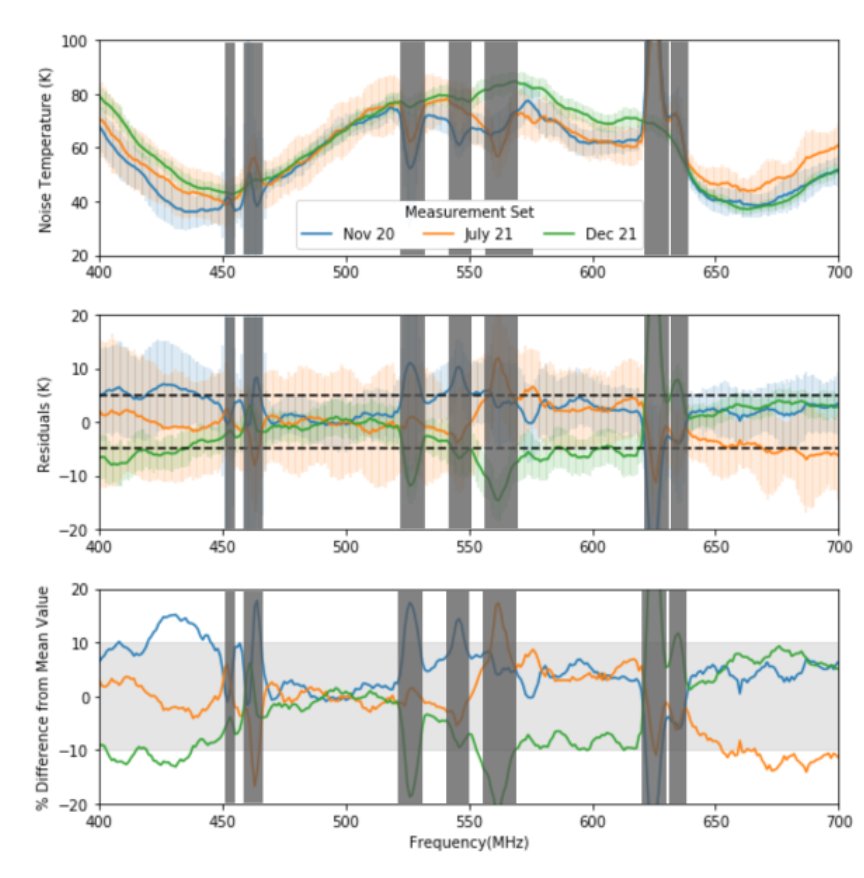}
\end{center}
\caption{The HIRAX antenna noise temperature is broadly consistent to within 5K/10\% over three multi-day measurement sets performed in November 2020, July 2021 and December 2021. The $\sim$60K average noise temperature is greater than the 30K allotted LNA noise budget, and this will be addressed in future designs. The shaded bands on each curve represent measurement uncertainty. Vertical gray lines mask RFI features. Measurements above 700MHz are contaminated by high levels of RFI, so we do not characterize this part of the HIRAX band.} 
\label{fig:yfacresults}
\end{figure}

We find the noise temperature of the HIRAX antenna to be frequency dependent, oscillating about $\sim$60K (Figure \ref{fig:yfacresults}), with a ripple of peak-to-peak height 40K. These measurements were repeated for four different feeds with two working polarizations. 

Figure \ref{fig:yfacresults} shows the noise temperature of HIRAX as computed during measurement sets performed in November 2020, July 2021 and December 2021. Each measurement set consists of multiple days. The individual curves on each plot represents a noise temperature for one polarization on one antenna as measured during a $<20$ minute window, which may include multiple hot and cold measurements averaged together. Each curve is plotted within a shaded band of the same color that is associated with the measurement statistical uncertainty (using standard error propagation for the Y-factor linear calculation, summing errors in measured $P_{\rm hot}$ and $P_{\rm cold}$ in quadrature). 
The results between measurement sets are broadly consistent within the measurement statistical uncertainty and within 10\% of the measurement mean (per frequency), excluding RFI-prone areas in the band (gray rectangles in Figure \ref{fig:yfacresults}). There are no discernible differences between the two feed polarizations.

We investigate relative measurements between different HIRAX feeds and different measurement days within a measurement set in addition to absolute noise temperature values. We find the HIRAX feed noise temperature results are repeatable across all three feeds and three separate measurement days to within $\pm$ 10\%.

\subsection{Reflection Assessment}
The frequency dependent features in both passive and active feed measurements are unexpected, and prompted the additional investigations described in this section. 
We consider that features may arise from a reflection in the system, which we investigated in two ways: (1) by raising the system lid, thereby increasing the cavity length and distance to nitrogen and (2) by lowering the liquid nitrogen level, thereby changing the distance between the antenna phase center and liquid nitrogen surface.

\begin{figure}[h!]
\begin{center}
\includegraphics[width=\textwidth]{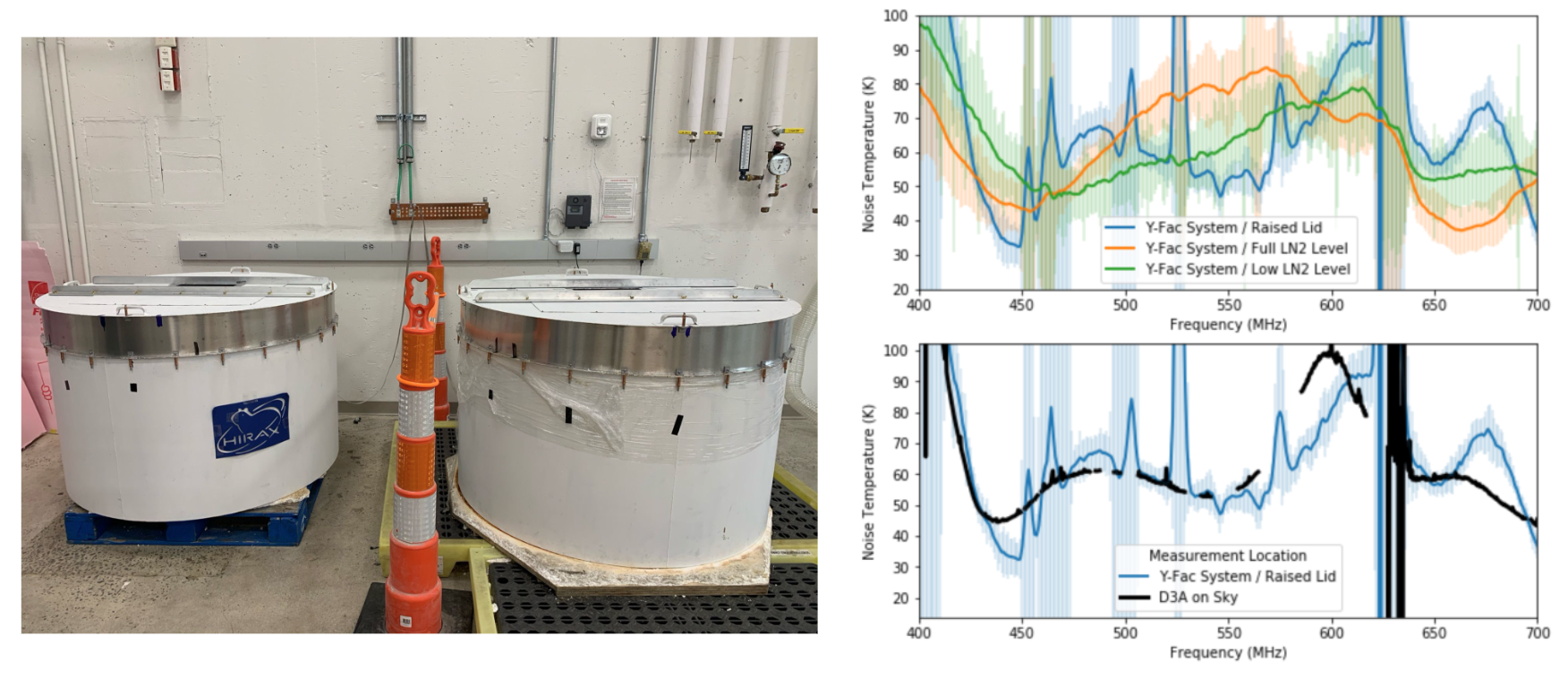}
\end{center}
\caption{Left: Cavity extenders of 15\,cm in length were constructed for the Y-factor system, in order to evaluate the impact of reflections and provide the capability to measure larger antennas. Right: (Top plot) Y-factor measurements of the active HIRAX antenna with both a raised lid and a lowered nitrogen level as compared with the full nitrogen level measurement. Shaded regions indicate measurement uncertainty. These measurements reveal a noise temperature profile that is overall consistent with original cavity measurements in level but differs in feature location. (Bottom plot) Y-factor measurements of the active HIRAX antenna (1) in the raised-lid system using room temperature and liquid nitrogen as hot/cold and (2) mounted to a 3m dish f/d = 0.23 dish, using on-sky measurements with sources in the beam and sources out of the beam as hot/cold.}
\label{fig:yfacreflections}
\end{figure}

To elongate the cavity, we constructed a cavity lid extender. This construction consisted of rolling 15\,cm wide aluminum pieces into the cavity circumference, and welding them together to form a ring. We then attached 24 L-brackets to the ring with pop rivets that mounted to a companion series of L-brackets bolted to the original cylinder. A system photo is in Figure \ref{fig:yfacreflections}. It is worth noting that in raising the lid, we are deviating from the system dimensions that were optimized for in simulations, and a predictable decrease in system performance is revealed in $S_{11}$ measurements

Y-factor measurements of the HIRAX antenna with both a raised lid and a lowered nitrogen level reveal a noise temperature profile that is overall consistent with original cavity measurements in level but differed in spectral structure (Figure \ref{fig:yfacreflections}, top right plot). These deviations can change the measured noise temperature at a given frequency by as much as 30K between the two cavity heights, and by up to 20K between the two nitrogen levels. Prior work anticipated a combined 11.5K maximum offset due to nitrogen surface reflections and cavity differences \cite{kuhn2021design} at the standard cavity height, which does not fully account for the measurement discrepancy seen at different nitrogen levels and requires renewed attention. Even so, given the consistent overall levels, the Y-factor measurement system may be used as a method of comparison between different antenna models prior to reflection mitigation.

The raised lid measurements were compared with on-sky Y-factor measurements taken by an active HIRAX feed mounted to a 3m dish belonging to the Deep Dish Development array (D3A)\cite{islam2020metrology} at the Dominion Radio Astronomy Observatory in Penticton, British Columbia, Canada. Both the Yale and D3A systems share a 75$\pm3$\,cm characteristic length scale between antenna phase center and metallic reflective surface (cavity bottom and dish parabola).
For the on-sky Y-factor, hot measurements were taken with a known source transiting overhead, and cold measurements were taken when no known sources were present in the beam. A comparison between the two measurements is shown in the bottom plot of Figure \ref{fig:yfacreflections}, revealing that features track in frequency location and amplitude across the full band.
The similarities between the two measurements suggest the presence of a reflection is modifying the measurement.  Conversely, measurements without raising the lid do vary from the D3A measurements, further suggesting that the spectral shape may originate from a specific configuration between the feed and reflective components in its beam. 

We remain encouraged by the close qualitative alignment between on-sky and lab-based measurements. Having a testbed with comparable performance to on-sky measurements will save considerable time and effort when evaluating future antenna models before deployment.

\section{Conclusion}
This paper describes two key characterization measurements performed for the HIRAX antennas, which yield antenna beam pattern and noise temperature results.

In beam measurements of the HIRAX antenna, we find that the antenna exhibits the expected gain and has a beam pattern in agreement with simulations. The design is shown to be repeatable, through comparisons of repeated beam and gain measurements across three different antennas. We additionally discovered an asymmetry in the co-polar beam, and notably high levels of cross-pol. Through comparisons with simulations and with measurements of similar antennas, we we identify this behavior to result from asymmetry in the way the antenna is fed, which will be corrected in future antenna versions. To complement HIRAX characterizations and provide a design comparison, we measured the CHIME antenna (of a similar design to HIRAX), finding a more symmetrized beam and lower cross-polarization levels. This comparison suggests that with minimal design modifications HIRAX can accomplish an improved antenna performance. This design is currently under assessment by the collaboration. 

We additionally designed and optimized a system to measure the noise temperature of the HIRAX feeds. Because HIRAX uses an embedded amplifier in the antenna, a system like the one we have developed is the only way to measure noise temperature in the lab. This measurement will be critical to verifying that the HIRAX feed design meets the noise specifications required to detect the faint cosmological signal of interest. 

This measurement system consists of two identical 4\,ft diameter cavities held at different temperatures to allow for a `Y-factor' measurement. The system design and early verification tests are outlined in Kuhn et al.~(2021)\cite{kuhn2021design}. The verification procedure utilized both passive and active HIRAX antennas to measure return loss and spectra of the two cavities and quantify systematic and statistical errors. As a secondary verification, Y-factor measurements were performed with
a passive feed and several previously-characterized amplifier chains (understood to $<$2K in most of the measurement band), to assess how accurately the test bed recovers a known quantity. Results indicate that this system is within 5 Kelvin tolerances with the main sources of error able to be removed or averaged down.

We perform the first measurements of the HIRAX antenna noise temperature, finding a $\sim$60K average noise temperature. The noise temperature profile is frequency dependent, oscillating about $\sim$60K, with a ripple of peak-to-peak height 40K that shifts when the cavity length is altered. These measurements were repeated for four different feeds with two working polarizations, and are shown to be consistent across antennas and measurement days. The $\sim$60K noise temperature is higher than the target of 30K, which motivates developing an upgraded antenna with lower noise. If an improved noise performance is not achieved, the measurement observing time can be increased to achieve the same result (though this would negatively impact survey speed). 

The Y-factor measurement system presented in this paper will be used to measure all 256 feeds used for the initial HIRAX deployment, as a spot check on production quality and consistency, and will also be used to evaluate future antenna prototypes.


\acknowledgments 
This work was made possible by the support from the Yale Wright Laboratory and Yale Center for Research Computing staff and administrators, and the Wright Laboratory computing and machine shop resources. In particular, we acknowledge Frank Lopez and Craig Miller for their help constructing the experiment and ensuring personnel safety. The beam measurements were made in the North Carolina State University Neofabrication Facility's anechoic chamber. We gratefully acknowledge discussions with Meiling Deng and Mark Halpern. This work benefited greatly from Meiling's CHIME feed design and simulations, and use of one of the CHIME feeds presented in this work. This work was supported by a NASA Space Technology Research Fellowship, and is based upon work supported by the National Science Foundation under Grant No. 1751763. KM acknowledges support from the National Research Foundation of South Africa. M.G.S. acknowledges support from the South African Radio Astronomy Observatory and National Research Foundation (Grant No. 84156). HIRAX is funded by the NRF of South Africa.

\bibliography{report} 
\bibliographystyle{spiebib} 

\end{document}